\documentclass[a4paper,12pt]{iopart}

\usepackage{graphicx,iopams}

\newcommand{\be}{\begin{equation}}
\newcommand{\ee}{\end{equation}}
\newcommand{\bq}{\begin{eqnarray}}
\newcommand{\eq}{\end{eqnarray}}

\begin{document}

\title{Solving the Anharmonic Oscillator: Tuning the Boundary Condition} 

\author{David Leonard and Paul Mansfield}
\address{Centre for Particle Theory, Durham University, Durham, DH1 3LE, UK}
\eads{\mailto{david.leonard@durham.ac.uk},\mailto{p.r.w.mansfield@durham.ac.uk}}

\begin{abstract}
We outline a remarkably efficient method for generating solutions to quantum anharmonic oscillators with an $x^{2M}$ potential. We solve the Schroedinger equation in terms of a free parameter which is then tuned to give the correct boundary condition by generating a power series expansion of the wavefunction in $x$ and applying a modified Borel resummation technique to obtain the large $x$ behaviour. The process allows us to calculate energy eigenvalues to an arbitrary level of accuracy. High degrees of precision are achieved even with modest computing power. Our technique extends to all levels of excitation and produces the correct solution to the double well oscillators even though they are dominated by non-perturbative effects.
\end{abstract}





\section{Introduction}
Harmonic oscillators are a corner-stone of many branches of physics. Consequently a large variety of methods have been used to study the eigenvalue properties of {\it anharmonic} 
oscillators (see \cite{Hioe:1978jj}\cite{kirsten} and references therein for a general review). High levels of accuracy have always been difficult to achieve due to slow convergence or often non-convergence of asymptotic perturbative expansions. For example the Bender Wu \cite{Bender:1969si} expansion of the quartic anharmonic oscillator ground state energy eigenvalue in positive powers of the coupling is known to be divergent for all non-zero values of the coupling. Methods of resumming asymptotic series \cite{Hardy} have been applied to generate approximate eigenvalues \cite{graffi}\cite{loeffel}. In addition some types of anharmonic oscillators are dominated by non-perturbative effects such as instantons \cite{Coleman:1978ae}. More innovative approaches have been required to produce a greater level of accuracy and account for these non-perturbative effects \cite{Plo}\cite{Castro}\cite{Buenda}\cite{Sanchez}. In addition to the numerical approaches some progress has been made in determining the analytic structure of certain anharmonic oscillators \cite{Singh:1977bk}\cite{Shifman:1988aj}. In particular \cite{Shifman:1988aj} outlines a type of anharmonic oscillator which is quasi exactly solvable with certain parts of the spectrum known exactly.

These problems extend into quantum field theory. For example, the renormalisation group implies that the energy eigenvalues in Yang-Mills theory cannot be solved for perturbatively. Strongly coupled theories in particular are hard to deal with using traditional techniques. Anharmonic oscillators are therefore of great interest because of their applicability in many branches of physics and because their mathematical properties often mirror those of other physical systems. 

We will outline a method of constructing solutions to the Schroedinger equation for an anharmonic oscillator of the form
\begin{eqnarray}
\label{gaho}
-\frac{d^2\Psi}{dx^2}+\rho x^2\Psi+gx^{2M}\Psi=E\Psi\\
\label{gPsibc}
\lim_{|x|\to\infty}\Psi=0
\end{eqnarray}
where $x$ is real and units are defined to absorb Plank's constant and the mass such that $\hbar=2m=1$. We do this initially by constructing a solution to the differential equation \eref{gaho} in terms of one free parameter for a given $\rho $ and $g$. We then vary this parameter until we observe the correct large $x$ behaviour determined by the boundary condition \eref{gPsibc} using a contour integral method of resummation. We find the energy eigenvalues with an arbitrary level of accuracy. The process is easily automated to produce very high levels of precision even with modest computing power.

In section \ref{tune} we will outline the basic method for the ground state of the quartic oscillator, $M=2$. In section \ref{excited} we will extend the method to produce excited wavefunctions and energy eigenvalues. Finally in section \ref{otherpots} we will show how this method can be extended to general anharmonic oscillators with an $x^{2M}$ potential as in \eref{gaho}. 

\section{Tuning for Large $x$}
\label{tune}
In this section we will find the ground state wavefunction and energy eigenvalues corresponding to the quartic anharmonic oscillator obtained from \eref{gaho} by setting $M=2$.
Since the ground state has no nodes we will construct the wavefunction in the form $\Psi=\exp(W)$. We will make an even powered $x$ expansion $W=\sum_{n=1}^\infty a_nx^{2n}$ since both the potential term and boundary condition are even. The coefficients $a_n$ are then determined in terms of the parameters $\rho $, $g$ and $E$ via \eref{gaho}. Having chosen two of these parameters the third must be determined by ensuring the correct boundary condition \eref{gPsibc}, which implies that $W\sim-\sqrt{g}x^3/3$ for large positive real $x$. Since our expansion for $W$ in positive powers of $x$ is only valid for small $x$ we shall resum by analytically continuing $x$ into the complex $s\equiv 1/x$ plane and using Cauchy's theorem to examine the large $x$ behaviour. We define
\begin{equation}
\label{L}
L(\lambda)=\frac{1}{2\pi i}\frac{1}{\lambda^3}\int_C ds\,\frac{e^{\lambda s}}{s}W(s).
\end{equation}
where $C$ is a large circular contour about the origin in the complex $s$ plane. The large $x$ asymptotic behaviour implied by the differential equation requires $W(s)$ to have a third order pole at the origin. This contributes a term $-\sqrt{g}/18$ to $L(\lambda)$ by Cauchy's theorem. When the boundary condition is satisfied we find that any remaining singularities of $W(s)$ lie to the left of the imaginary axis. The contribution from these is exponentially suppressed in $L(\lambda)$ so that in the large $\lambda$ limit only the singular contribution at the origin remains, $\lim_{\lambda\to\infty}L(\lambda)=-\sqrt{g}/18$. In reality  \eref{L} is not calculated exactly but by truncating $W$ at some order $x^{2N}$. Thus
\begin{equation}
L_N(\lambda)\equiv \frac{1}{2\pi i}\frac{1}{\lambda^3}\int_C ds\,\frac{e^{\lambda s}}{s}\sum_{n=1}^N\frac{a_n}{s^{2n}}=\sum_{n=1}^Na_n\frac{\lambda^{2n-3}}{\Gamma(2n+1)}\approx L(\lambda)
\end{equation}
where in the evaluation of the contour integral we used the identity $\int_C ds\,s^{-n}\exp(\lambda s)=2\pi i\lambda^{n-1}/\Gamma(n)$ for $n<0$ \cite{Magnus}.

We proceed by finding our $x$ expansion in $W$ and look for the correct behaviour in $L_N(\lambda)$. We will consider solutions with a fixed coupling $g=1$ and look at the relationship between $E$ and $\rho $. We do this without loss of generality since our parameters are related by scaling properties of the Hamiltonian, as first noted by Symanzik and discussed in \cite{Simon:1970mc}. To help us we will scale $x\to cx$ ($c\in\mathbb{R}$) in the differential equation \eref{gaho} in such a way that we are free to place a restriction on our expansion for $W$. We can choose $k\equiv a_1/a_2$ at least up to a sign, say $k=\pm 4$. Now substituting $W$ into our scaled differential equation and comparing coefficients of $x^{2n}$
\begin{eqnarray}
Ec^2=-2a_1\label{E}\\
\rho c^4=4a_1^2+12a_2\label{m}\\
c^6=16a_1a_2+30a_3\,.\label{g} 
\end{eqnarray}
We eliminate $c$ to find expressions for $E$ and $\rho $ in terms of $a_2$ and $a_3$
\begin{eqnarray}
\label{Em}
E=\frac{-2ka_2}{(16ka_2^2+30a_3)^{\frac{1}{3}}}\\
\rho =\frac{4k^2a_2^2+12a_2}{(16ka_2^2+30a_3)^{\frac{2}{3}}},
\end{eqnarray}
whilst for $n\ge3$ we have 
\begin{equation}
\label{recur}
a_{n+1}=-\left(\sum_{m=1}^n4m(n-m+1)a_ma_{n-m+1}\right)/\left(2(n+1)(2n+1)\right)\,,
\end{equation}
giving $a_{n+1}$ in terms of $a_2$ and $a_3$.

Our goal is now to determine $a_3$ for a given $a_2$ in such a way that the boundary condition is satisfied. We do this by tuning $a_3$ until the correct large $\lambda$ behaviour is observed in $L_N(\lambda)$. To illustrate the process we shall choose positive $k$, $k=4$ and $a_2=-3/16$. With this sign choice and $a_2$ we get a zero $\rho$ term. We choose a fairly modest $N$ initially, guess a value of $a_3$ then plot $L_N(\lambda)$ and $L_{N-1}(\lambda)$. $L_N$ and $L_{N-1}$ only provide a good approximation to $L(\lambda)$ for values of $\lambda$ up to the point where they appreciably diverge from each other. Therefore we restrict our consideration of $\lambda$ to within this range. 

\begin{figure}
$\begin{array}{ccccc}
\includegraphics[scale=0.25]{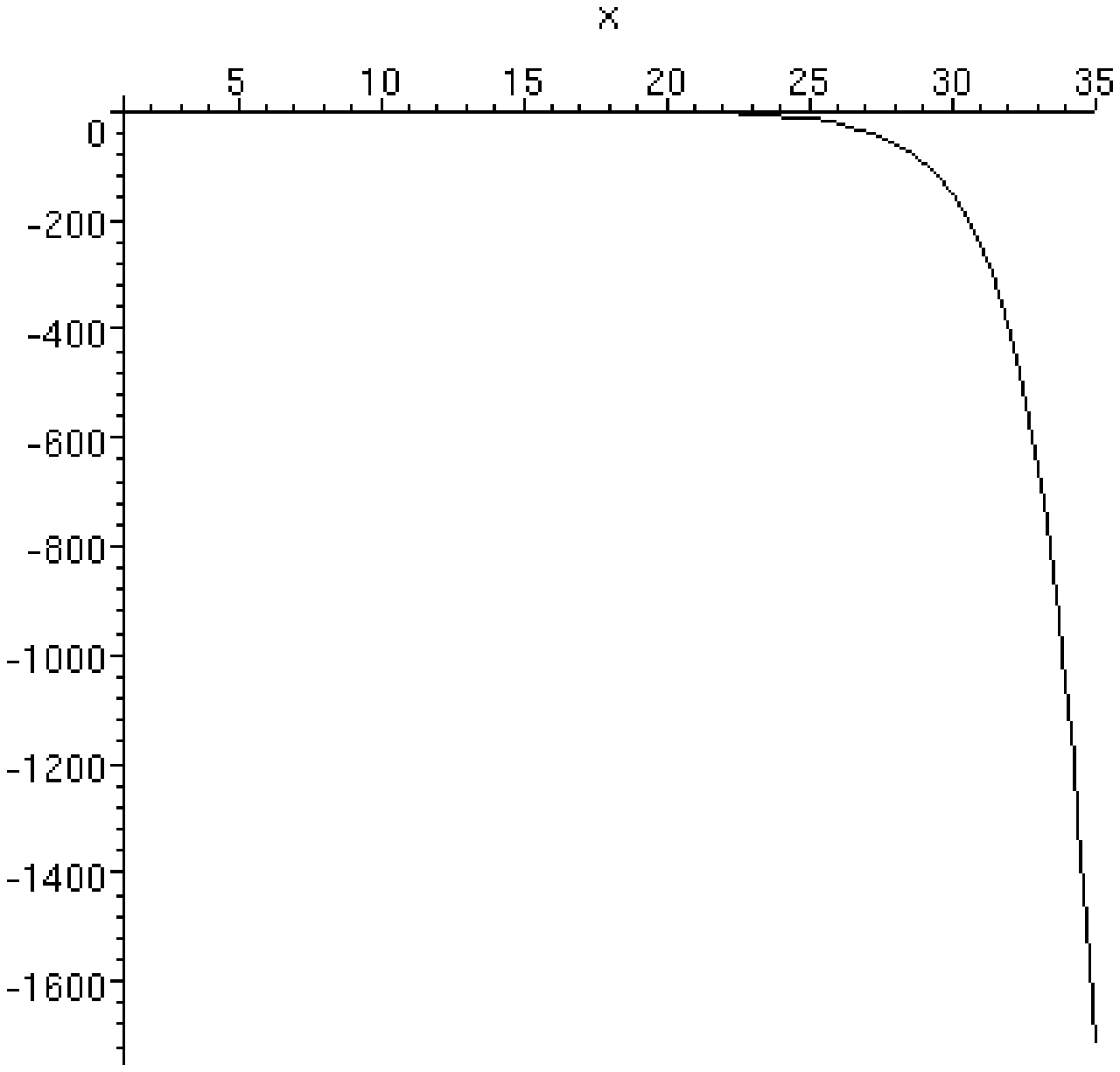}&\hspace{5pt}&\includegraphics[scale=0.25]{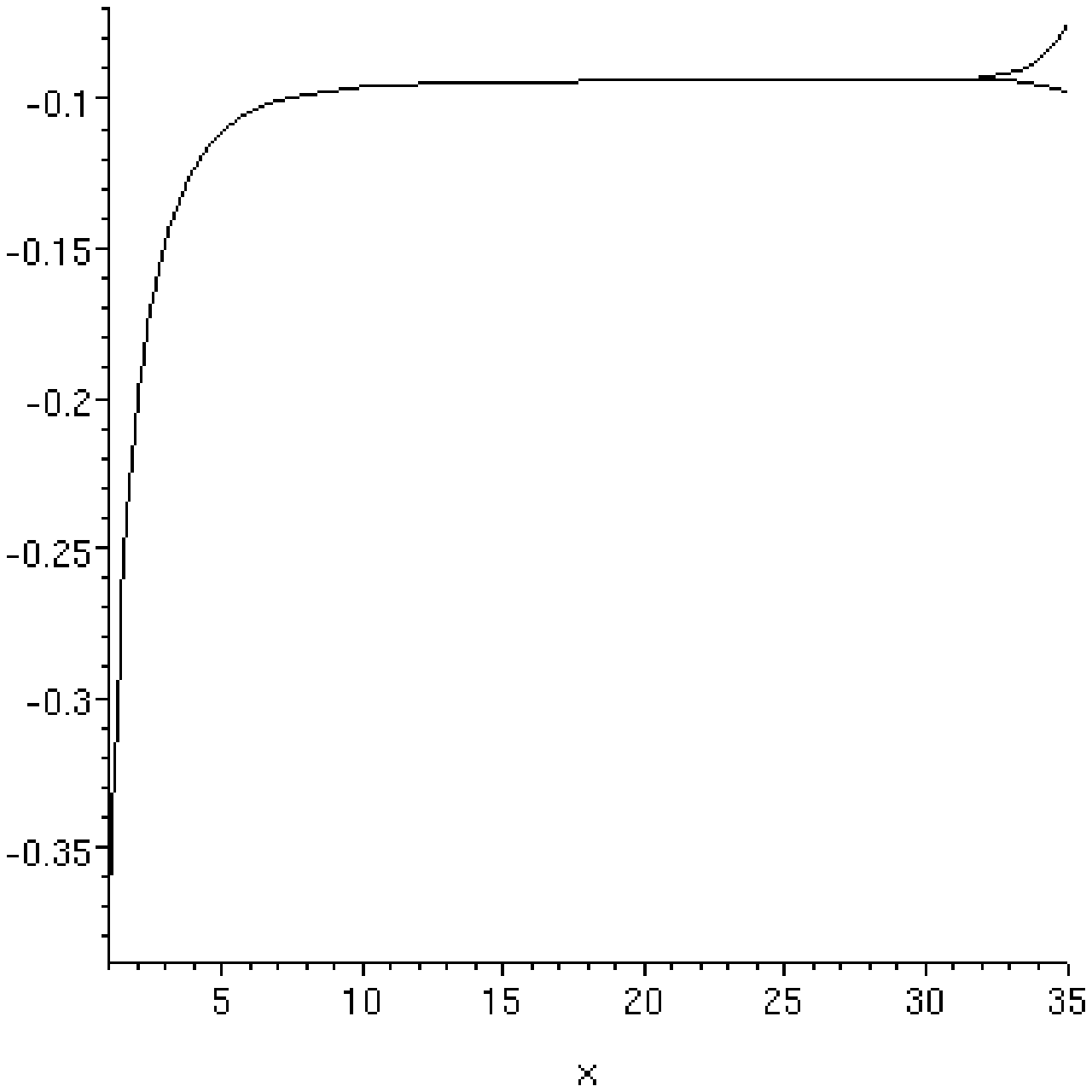}&\hspace{5pt}&\includegraphics[scale=0.25]{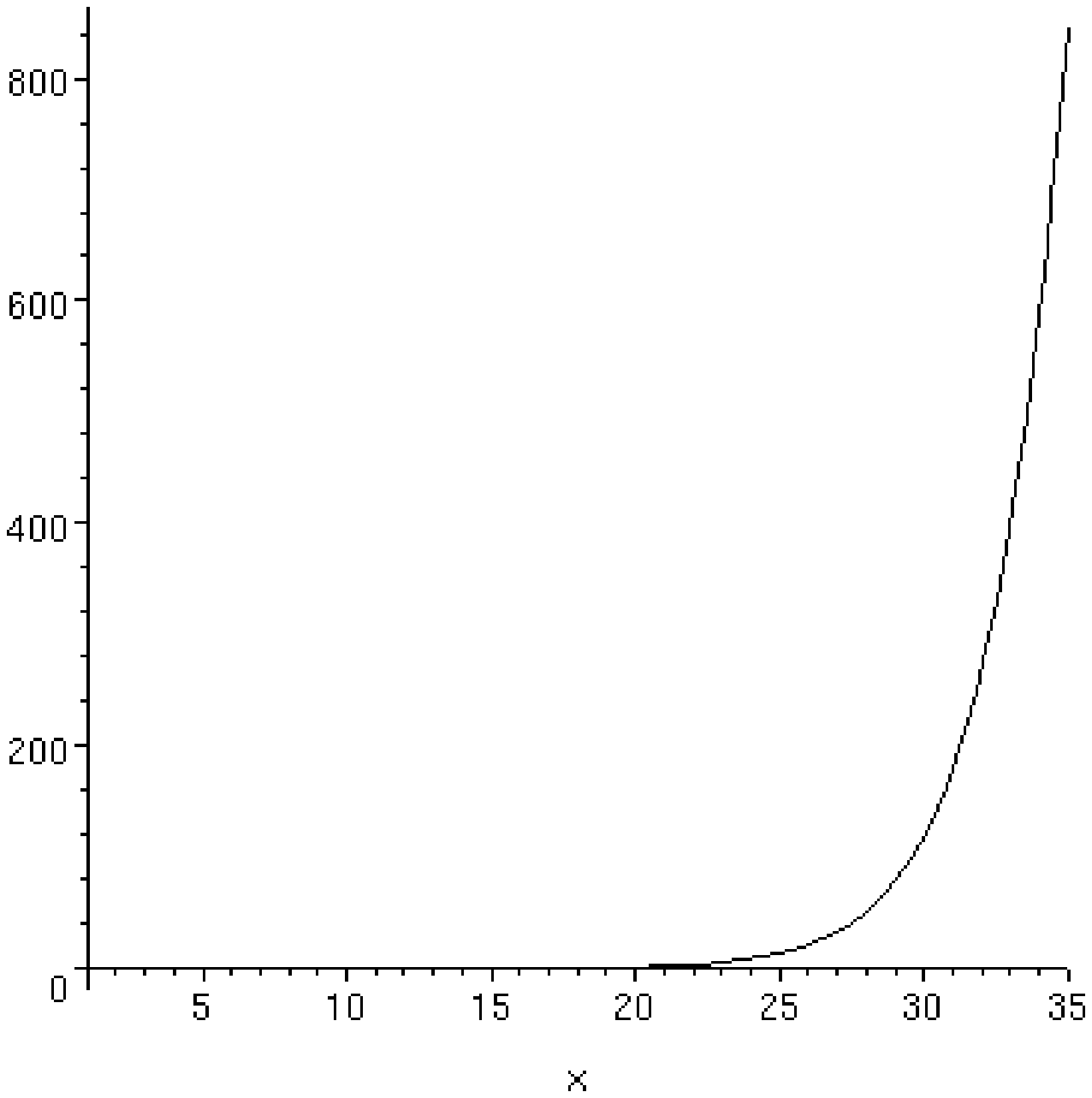}\\
$a) $a_3=0.015$ - too small$&&$b) $a_3=0.0193604$ - optimal$&&$c) $a_3=0.025$ - too big$
\end{array}$
\caption{$L_N(\lambda)$ with $N=19,20$ for $a_2=-3/16$}
\label{extune}
\end{figure}

With $a_3$ too small we encounter a curve rapidly decreasing such as in figure \ref{extune}a. With $a_3$ too large we encounter a curve rapidly increasing as in figure \ref{extune}c. An optimal value of $a_3$ will give a curve flattening as we increase $\lambda$ as in \ref{extune}b. We tune $a_3$ until we achieve this. As $a_3$ gets closer to its correct value the exponential behaviour in figures \ref{extune}a and \ref{extune}c becomes less pronounced within our range of acceptable $\lambda$ and flatness becomes a less well defined concept. This determines our level of accuracy for determining $a_3$. 
To improve our accuracy we must increase $N$ in order to consider larger $\lambda$. As we consider these larger $\lambda$ we again encounter the exponentially increasing / decreasing behaviour which enables us to further tune $a_3$ to a greater accuracy. 

We completed this procedure in this zero $\rho$ case and determined $a_3$ to 6 significant figures with $N=20$. With $N=100$ we tune $a_3$ to 30 significant figures and with $N=300$ we get
\be
\fl a_3=0.01936043720245950419201997531721233596425589581549397570027615152,
\ee
to 65 significant figures and we find $L(\lambda)\approx-0.0934774$ which is remarkably close to the predicted value of $-0.0934723$. This calculated figure of $a_3$ is accurate to the stated number of digits i.e. 65 significant figures and in agreement with existing literature \cite{Plo}\cite{Castro}\cite{Buenda}\cite{Sanchez} at least up to the 10-16 significant figures they quote. With $a_3$ determined we calculate the ground state energy via \eref{Em}
\be
\fl E_0=1.0603620904841828996470460166926635455152087285289779332162452417
\ee
again quoted accurately up to 65 significant figures. 

Calculating large numbers of terms is easy, even with modest computing power, given the linear nature of the calculations. The tuning process is easily automated. 
\begin{figure}
\label{hbargraphs}
$\begin{array}{ccccc}
\includegraphics[scale=0.33]{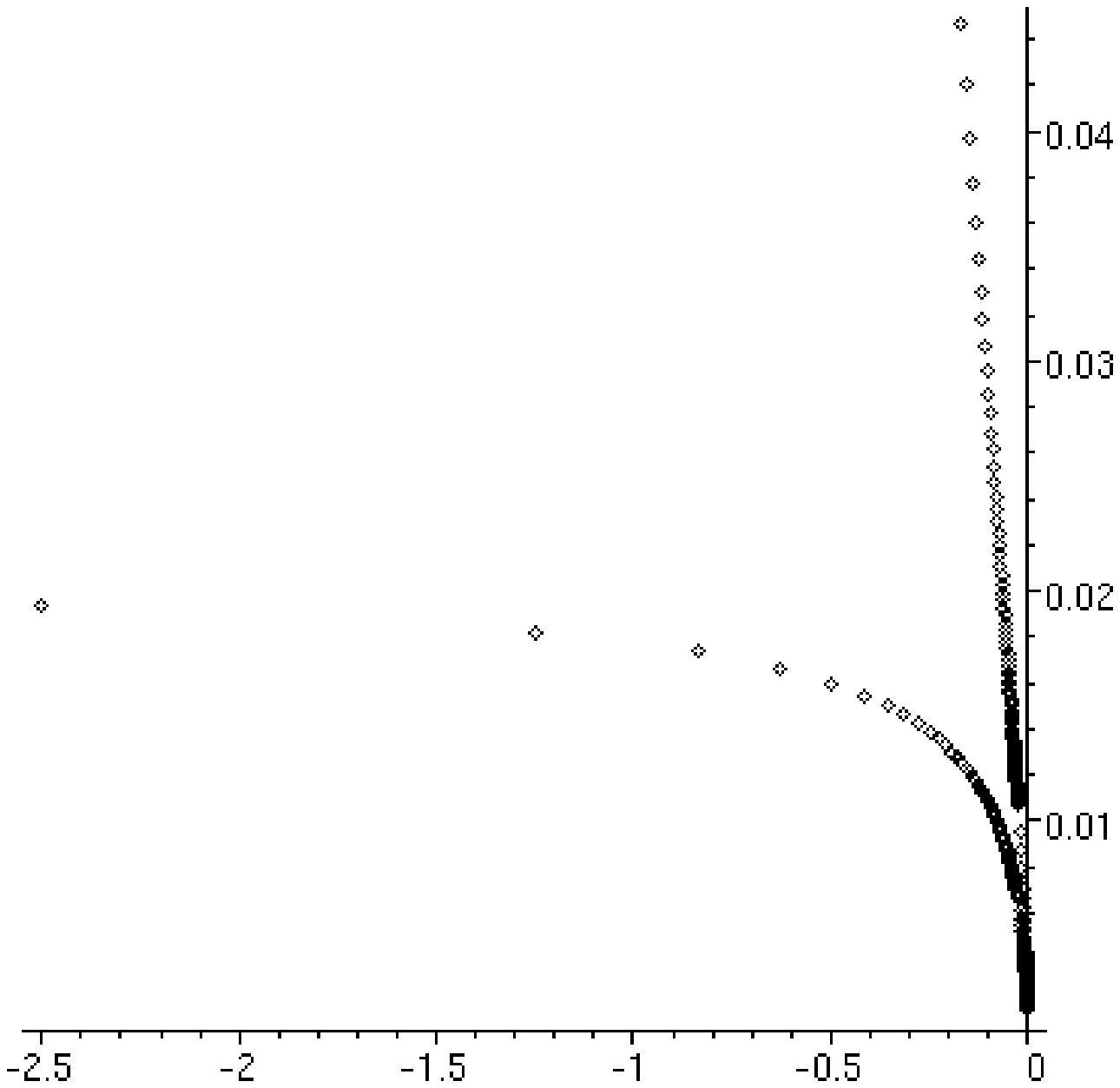}&\hspace{5pt}&\includegraphics[scale=0.33]{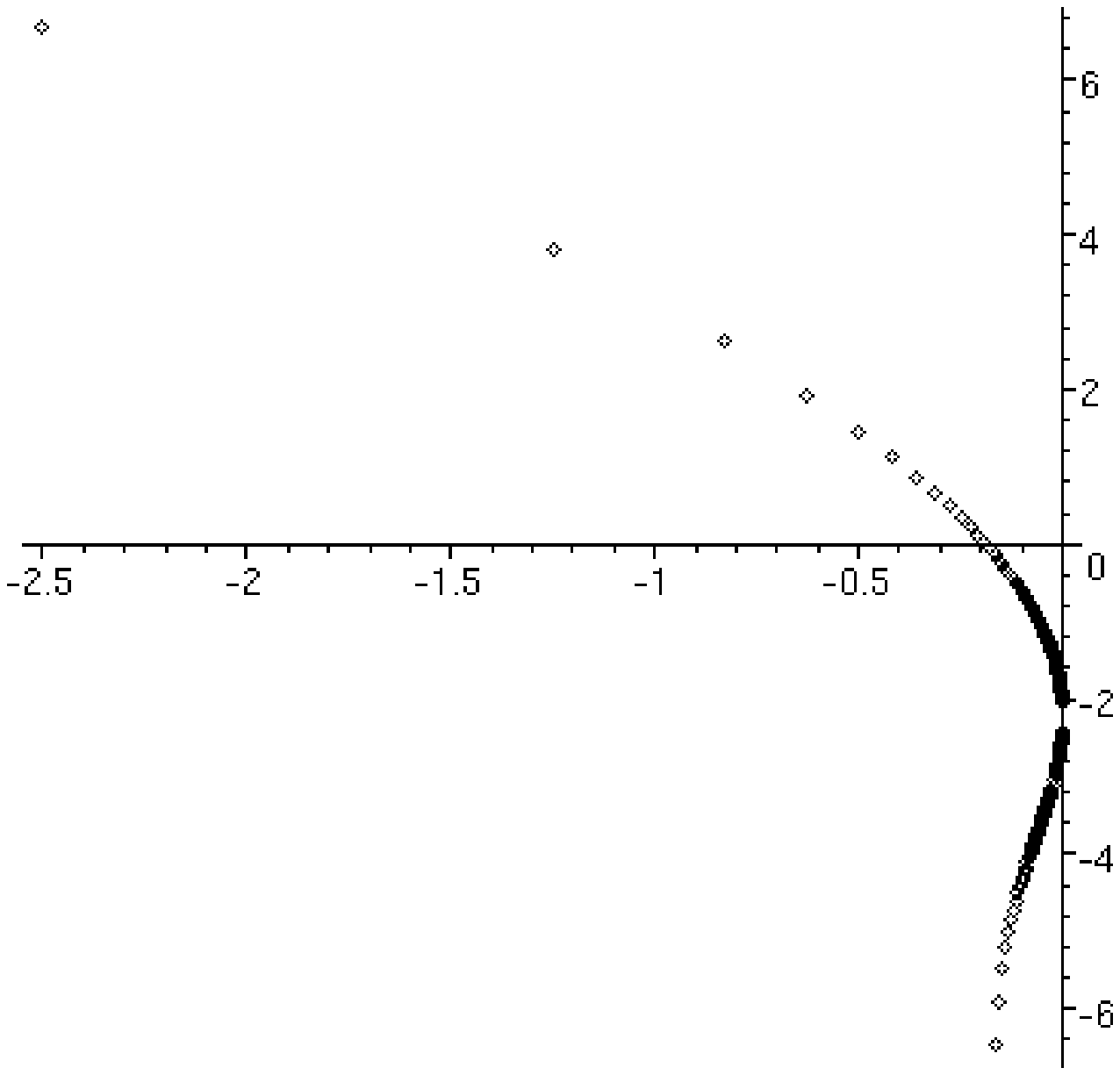}&\hspace{5pt}&\includegraphics[scale=0.33]{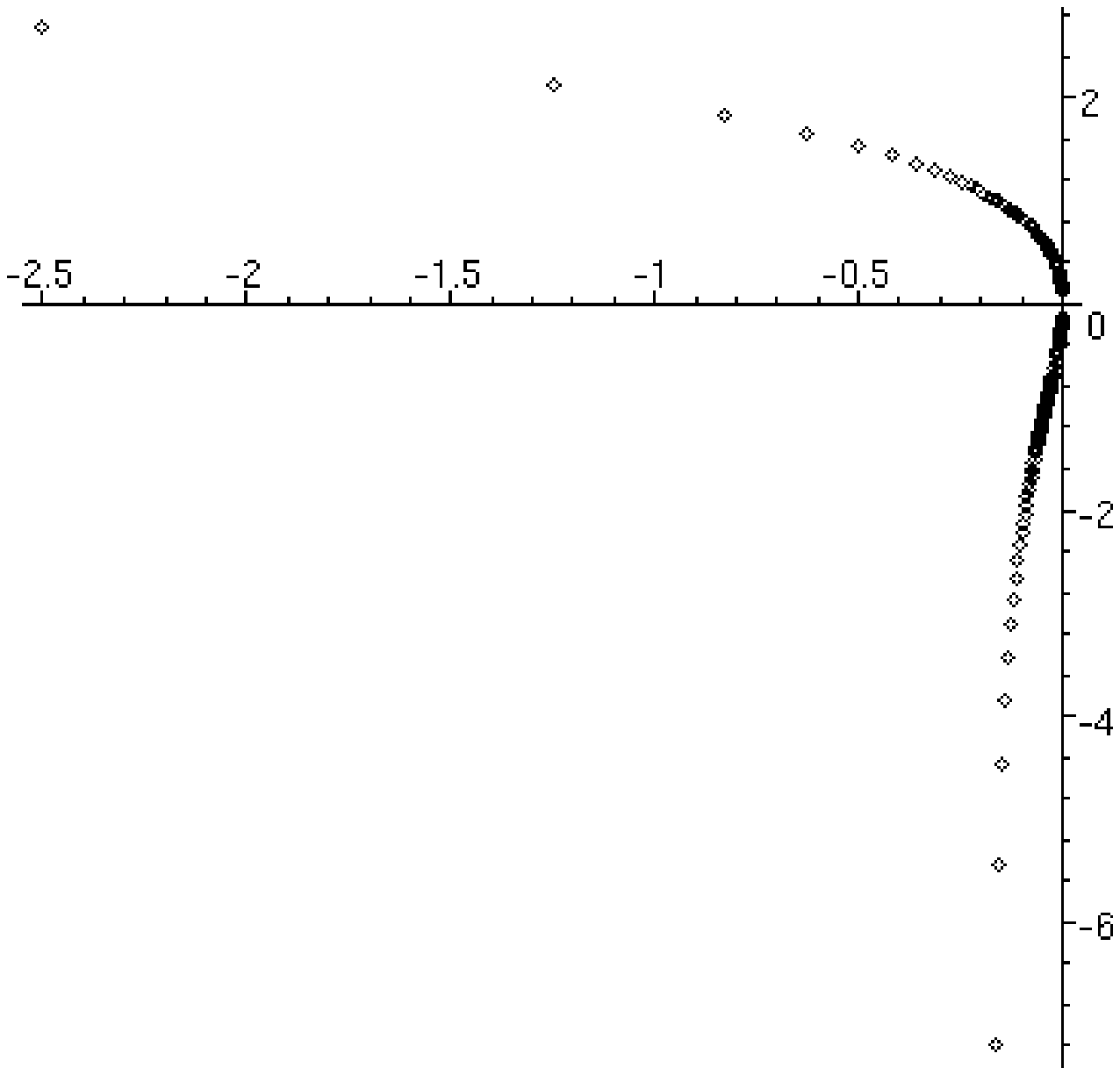} \\
$a) $a_3$ as a function of $a_2$ $&&$b) $\rho $ as a function of $a_2$ $&&$c) $E$ as a function of $a_2
\vspace{7pt}
\end{array}$
\vspace{-7pt}\caption{$\rho $, $E$ and $a_3$ as functions of $a_2$.}
\end{figure}

\begin{figure}
\begin{center}\includegraphics[scale=0.5]{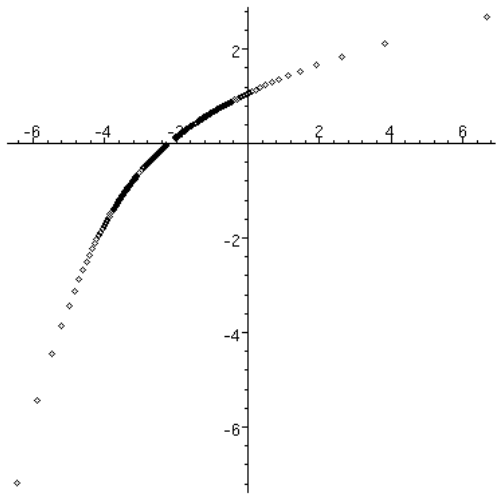}\end{center}
\caption{$E$ as a function of $\rho $}
\label{Em2}
\end{figure}

We repeat this for various $a_2$ and plot the results in figures \ref{hbargraphs} and \ref{Em2}. The two branches correspond to differing sign choices of $k$. With $k=+4$ we found solutions corresponding to positive energy. The solutions have a positive $\rho $ term for $a_2<-3/16$ and a negative term for $-3/16<a_2\le 0$. With $k=-4$ and $0>a_2>-3/16$ we found negative energy eigenvalues corresponding to $\rho 	<0$.

We also verify that non-zero $\rho$ terms correspond to the literature by for example calculating the $\rho =-1$, $g=1$ energy eigenvalue. In doing so we must tune $a_2$ with $k=+4$ to obtain the correct $\rho$ term. We found that $a_2=0.004048768355681543705$ approximated $\rho =-1$ with an error in the order of $10^{-16}$. This value of $a_2$ is within $\pm 5^{-21}$ of the correct $a_2$ required to evaluate $\rho $ exactly. The energy eigenvalue produced from this approximate value of $a_2$ gave us the same eigenvalue as stated in previous literature to within the 16 significant figures available for comparison. This is an example of an eigenvalue where instanton effects would normally dominate and perturbative techniques in $\hbar$ or $g$ would fail.

We now explain why this method of tuning is so sensitive. With $M=2$ the differential equation \eref{gaho} without the boundary condition \eref{gPsibc} in general has an asymptotic large positive $x$ solution of the form
\begin{equation}
\label{asymx}
\Psi_l=\exp\left(-\frac{\sqrt{g}}{3}x^3\right)+A\exp\left(\frac{\sqrt{g}}{3}x^3\right).
\end{equation}
For $A<0$, $\Psi_l$ has zeros along the real $x$ axis however for $A>0$, $\Psi_l$ has zeros in the complex $x$ plane off the real axis. Our boundary condition \eref{gPsibc} requires us to take $A=0$ in which case $\Psi_l$ has no zeros. We note that for $A\ne 0$, $\log\Psi_l$ will have a pole (possibly part of a cut). Such a pole contribution in the right half $x$ plane would spoil our resummation of the large $x$ behaviour. We have numerically determined the location of zeros in our wavefunctions for varying $a_3$ and shown that they numerically approximate the location of the zeros in our asymptotic large $x$ solution for varying $A$. Thus varying $a_3$ corresponds to varying $A$ in \eref{asymx}. The presence of these poles is responsible for the rapidly increasing /decreasing behaviour for values of $a_3$ on either side of the correct one due to the exponential factor in \eref{L}. It is this behaviour that allows us to select the correct value of $a_3$ to any specified level of accuracy.

\section{Excited States}
\label{excited}
We now construct the excited states and energy eigenvalues of the quartic anharmonic oscillator. 
Firstly we write the $q$th excited state as $\Psi_q=P_q\Psi_0$ where the energy is $E=E_q+E_0$, and 
$\Psi_0$ is the ground state obtained in the previous section.
For $q$ odd $P_q$ is odd and for $q$ even $P_q$ is even. We therefore expand $P=\sum_{n=0}^\infty c_nx^n$ and sum only over even or odd values of $n$ as appropriate. We set either $c_0$ or $c_1$ to unity as a choice of normalisation. The remaining $c_n$ and $E_q$ are then solved for using a recurrence relation in terms of either $c_2$ or $c_3$. This is easily obtained from our new differential equation which comes from substituting $\Psi_q$ into \eref{gaho} to obtain
\begin{equation}
\label{Peqn}
\frac{d^2P}{dx^2}+2\frac{dW}{dx}\frac{dP_q}{dx}+E_qP_q=0.
\end{equation}
This differential equation has two types of large $x$ solution. Either
\begin{equation}
\label{largeP}
P\sim \exp{\left(-\frac{E_q}{2\sqrt{g}x}\right)}\qquad\mbox{or}\qquad P\sim \exp{\left(2\sqrt{g}x^3\right)}.
\end{equation}
For the correct boundary condition \eref{gPsibc} we must choose the first type of solution. We therefore construct
\begin{equation}
T_N(\lambda)\equiv \frac{1}{2\pi i}\int_C ds\,\frac{e^{\lambda s}}{s}\sum_{n=0}^N\frac{c_n}{s^{2n\alpha}}=\sum_{n=0}^Nc_n\frac{\lambda^{2n\alpha}}{\Gamma(2n\alpha+1)}
\end{equation}
and look for a flat curve as we tune $c_2$ or $c_3$. We have introduced an additional parameter $\alpha$ by substituting $s\to s^\alpha$ in $P(s)$ since we find that $P(s)$ has a more limited region of analyticity than $W(s)$ when the boundary condition is satisfied. Here we only assume that $P(s)$ is analytic in some wedge shaped region radiating from the origin and containing the real axis. Singularities outside of this region of analyticity are observed in $T_N(\lambda)$ in the form of oscillations. They can however be rotated in the complex $s$ plane so that they lie to the left of the imaginary axis by reducing the parameter $\alpha<1$. Having done this the singularities become exponentially suppressed.

\begin{figure}
$\begin{array}{ccccc}
\includegraphics[scale=0.33]{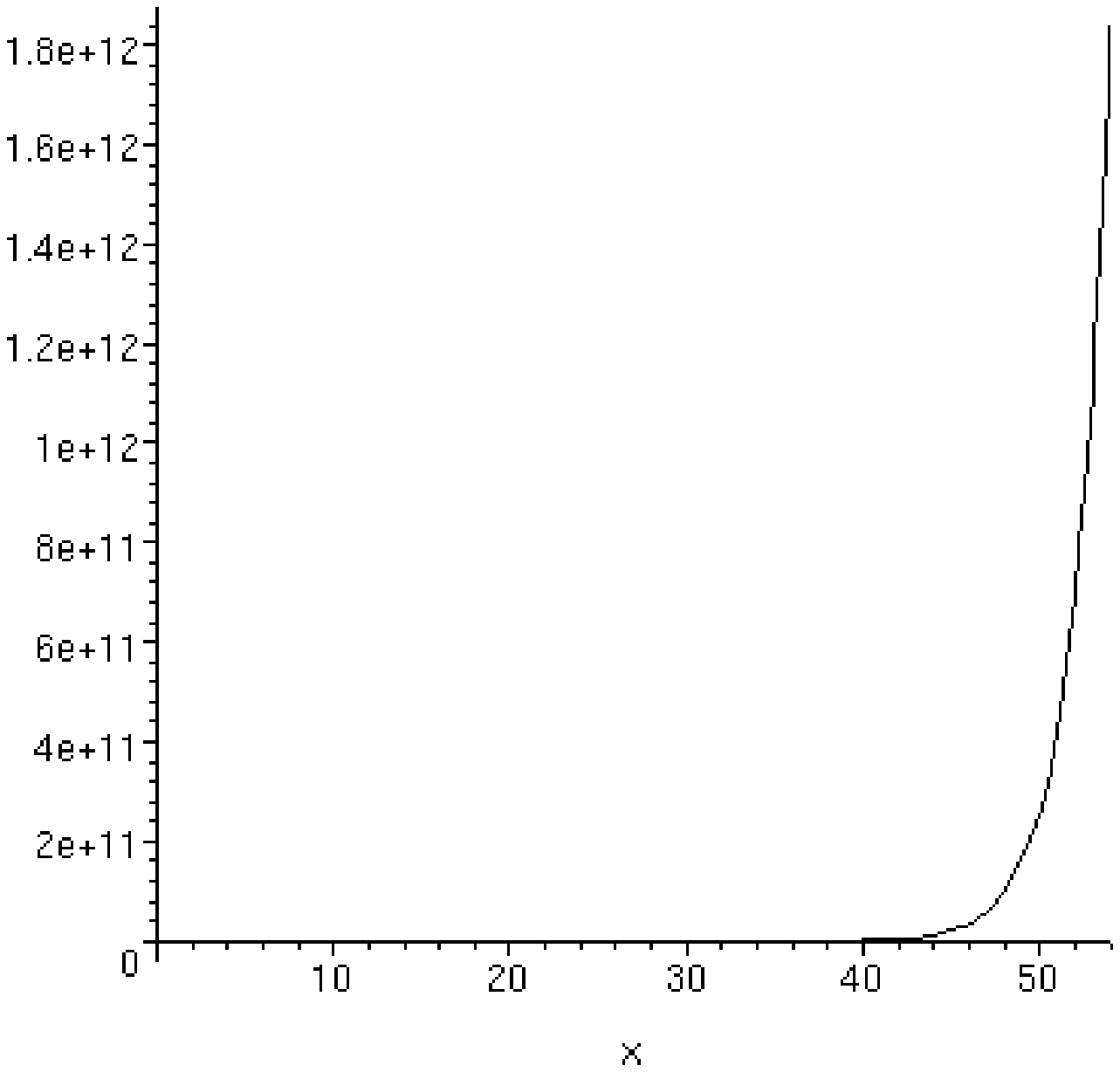}&\hspace{7pt}&\includegraphics[scale=0.33]{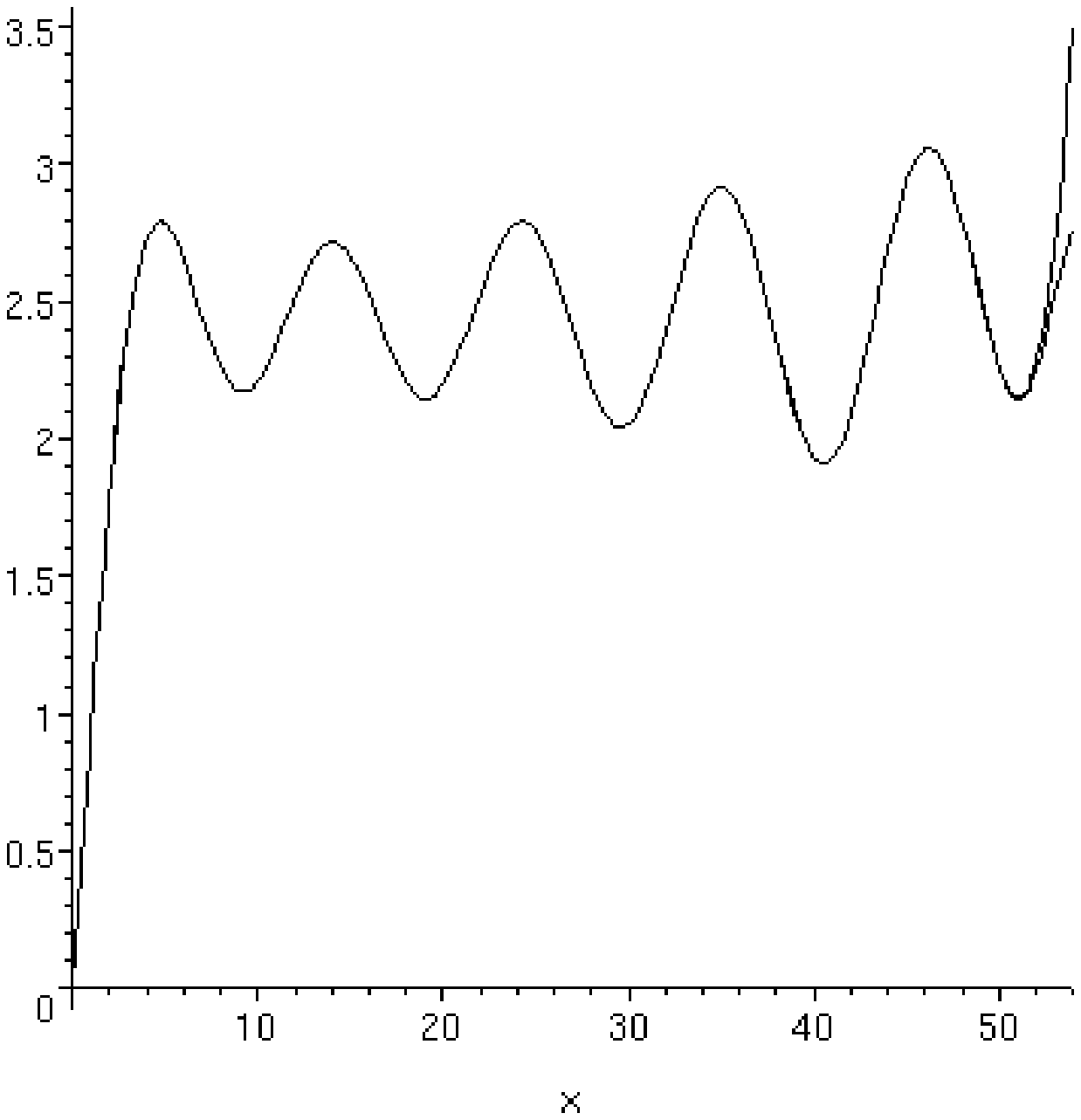}&\hspace{7pt}&\includegraphics[scale=0.33]{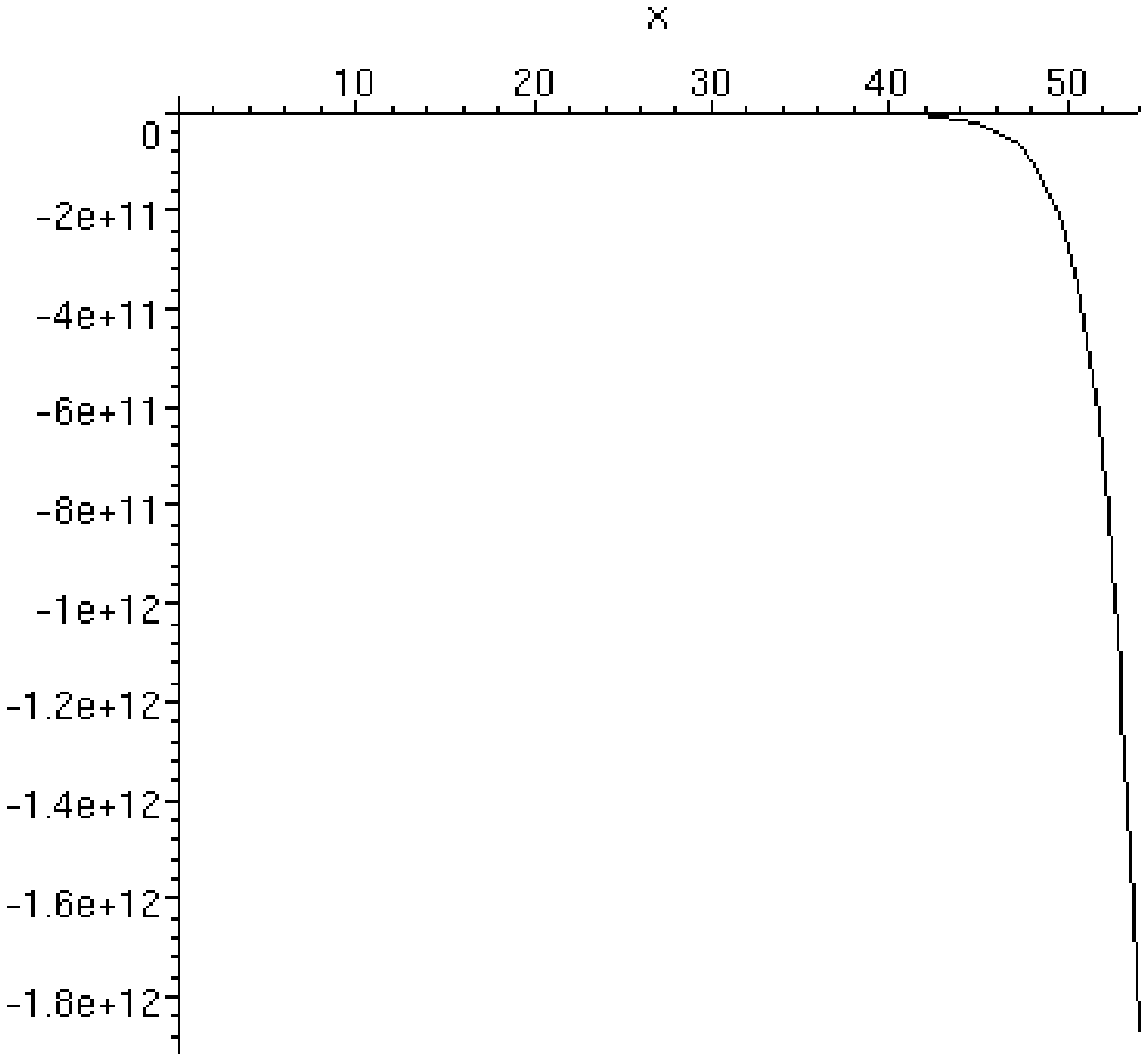} \\
$a) $\tau=0.1&&$b) $\tau\approx \tau_1\approx 0.14584&&$c) $\tau=0.2\\
\includegraphics[scale=0.33]{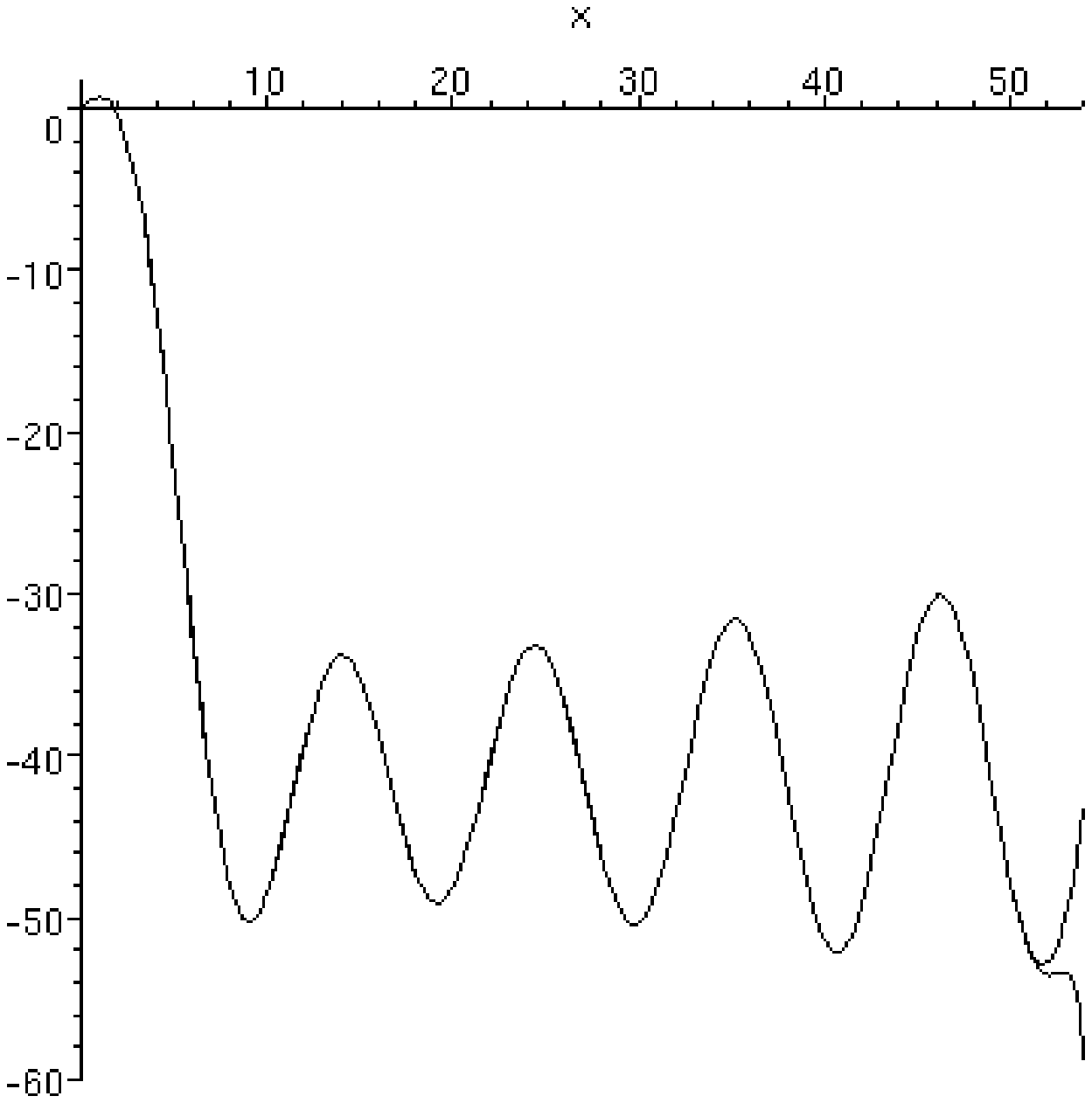}&\hspace{7pt}&\includegraphics[scale=0.33]{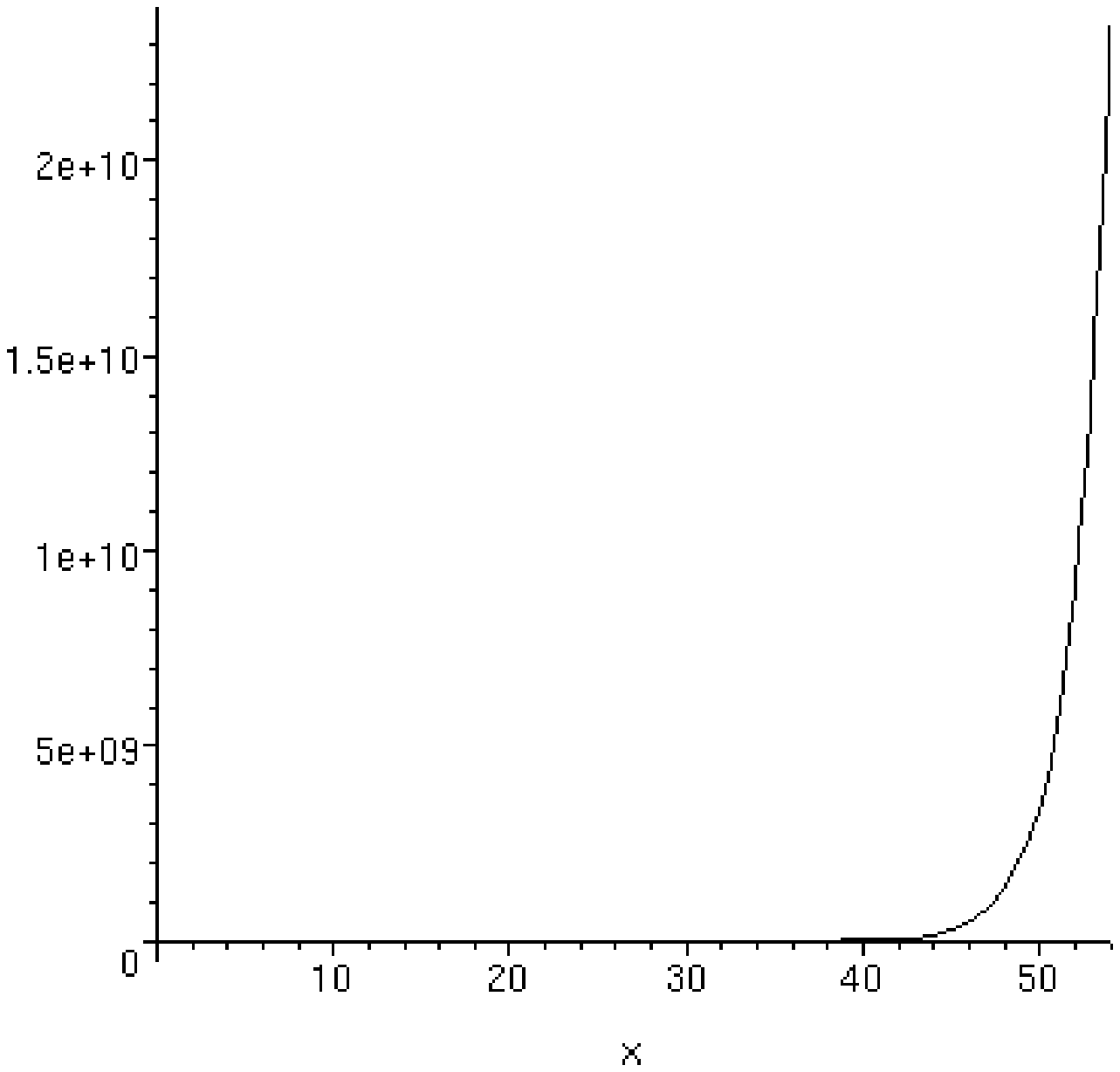}&\hspace{7pt}&\hspace{7pt}\\
$d) $\tau\approx \tau_2\approx 1.99546&&$e) $\tau=2&&\\
\end{array}$
\vspace{-7pt}\caption{$T(\lambda)$ with an odd prefactor}
\label{Tc3}
\end{figure}

We illustrate the process in the zero $\rho$ case for the odd eigenfunctions. There will be multiple values of $\tau\equiv -c_3$ that correspond to different levels of odd excitation. Let us label these $\tau_n$ in such a way that $\tau_{n+1}>\tau_n$. With $\tau<\tau_1$ we obtain a rapidly increasing curve however with $\tau_1<\tau<\tau_2$ we get a rapidly decreasing curve (figure \ref{Tc3}). We follow our tuning procedure in the same manner as for the ground state however this time we do not encounter a flat curve but oscillations. These result from a pole or cut outside of our region of analyticity. We could however take a smaller $\alpha$ to recover a flat curve and proceed with our tuning procedure. For $\tau=\tau_1$ we found  $\alpha=0.6$ sufficient to achieve this.

We can produce the full spectrum of eigenvalues by continuing to vary $\tau$. We find that as $\tau$ passes through a value $\tau_n$ we switch from the rapidly growing to rapidly decreasing behaviour. With $\tau_3>\tau>\tau_2$ for example we switch back to the rapidly increasing curve. This alternating behaviour continues with higher excitations as illustrated in figure \ref{Tc3}. Exactly the same procedure works for even excitations but we vary $c_2$ instead of $c_3$. Having found an eigenstate through this method we cannot immediately tell which energy level it corresponds to. To do this we could plot the prefactor using a similar contour integral method of resummation. We then count the number of nodes. We did this for some of the lower excitations. We calculated excited states up to $q=39$ with $g=1$ and again found exact agreement to the quoted level of accuracy in previous literature \cite{Plo}\cite{Castro}\cite{Buenda}\cite{Sanchez}. We give some of these eigenvalues in the appendix. 

Whilst we cannot attribute the rapidly increasing / decreasing behaviour of $T_N(\lambda)$ to zeros in $P_q$ we believe that a similar effect is encountered this time due to the large $x$ behaviour. There were two types of large $x$ behaviour \eref{largeP} that we were able to derive from the differential equation \eref{Peqn}. We chose the first in order to satisfy our boundary condition \eref{gPsibc}. When $c_2$ or $c_3$ do not correspond to an energy eigenstate we believe that we are obtaining the second type of solution. We note that such large $x$ behaviour would give an additional pole contribution to $T(\lambda)$. Again our resummation is conveniently spoilt. We have numerically verified this result by plotting $P$ for large real values of $x$ for a range of $c_3$ by exploiting Cauchy's theorem.

\section{Other Potentials}
\label{otherpots}
In this section we consider other values of $M$ in \ref{gaho}. The large positive $x$ behaviour is now $W\sim-\sqrt{g}x^{M+1}/(M+1)$. We should therefore redefine our $L_N(\lambda)$ for a general $x^{2M}$ potential
\begin{equation}
L_N^M(\lambda)\equiv \frac{1}{2\pi i}\frac{1}{\lambda^{(M+1)\alpha}}\int_C ds\,\frac{e^{\lambda s}}{s}\sum_{n=1}^N\frac{a_n}{s^{2n\alpha}}=\sum_{n=1}^Na_n\frac{\lambda^{(2n-M-1)\alpha}}{\Gamma(2n\alpha+1)}.
\end{equation}
where again we introduce the parameter $\alpha$ since for $M>2$ we find that $W(s)$ is analytic within a more limited region. Our prescription of reducing $\alpha<1$ will therefore be required to rotate these singularities to the left of the imaginary axis where they become exponentially suppressed.

Our $a_n$ are again determined via the differential equation in the same manner as before. We apply the rescaling $x\to cx$ so that we can fix $a_1/a_2=\pm4$ as before. We pick a value of $a_2$ and use  \eref{recur} to solve for all of the coefficients in terms of $a_{n+1}$. This relation now holds for $n\ge2$ but not $n=M$. In its place we have
\begin{equation}
c^6g=2(M+1)(2M+1)a_{M+1}+\sum_{n=1}^M 4n(M-n+1)a_Ma_{n-M+1}
\end{equation}
which is then substituted into \eref{E} and \eref{m} to give $E$ and $\rho $.

Our procedure is now the same as for the $M=2$ case. We do find however that for increasing $M$ the region of analyticity becomes smaller and therefore an increasingly small $\alpha$ is required. We performed this procedure with $M$ ranging from $2$ to $50$ for $g=1$ and found results matching those in \cite{Castro} for $M=2,3,4$. Having determined the ground state we have applied the technique outlined in section \ref{excited} to obtain some excited energy eigenvalues. Again these are in complete agreement with \cite{Castro}.  

\section{Summary}
We have developed a method for calculating the relationship between the physical parameters of a general $x^{2M}$ anharmonic oscillator. The equations we solve are linear and the process of refining our estimate is easily automated. We can calculate the physical quantities and wavefunctions for all levels of excitation to an arbitrary level of accuracy with an error that can be reduced by increasing the number of terms in our expansion. Using modest computing power we have demonstrated that high degrees of accuracy can be obtained very quickly. Our technique overcomes some of the deficiencies of traditional perturbative techniques
which rely on coupling constant expansions and so do not immediately reveal the effects of instantons, for example. Finally, we note that the analytic continuation of quantum mechanical systems into complex configuration space has recently been studied in $\mathcal{PT}$ symmetric quantum mechanics (see \cite{Bender:2005tb} and references therein). We believe that understanding the properties of Hermitian theories in the complex plane is still of great interest.   

Finally we note that in the case of the quasi exactly solvable solutions studied in \cite{Shifman:1988aj}, the expansion of both $W$ and $P$ in powers of $x$ becomes truncated. In this type of solution it is more obvious that the correct boundary condition is satisfied by the large $x$ behaviour. This is trivially reflected in our resummation technique. We have numerically verified that the results of \cite{Shifman:1988aj} are correctly reproduced for some specifc choices of an $x^6$ polynomial potential.

\appendix
\section*{Appendix}
Below we give some of the excited energy eigenvalues for the $\rho=0$ quartic ($M=2$) anharmonic oscillator. The results represent accurate eigenvalues rounded to 48 significant figures.
\begin{indented}
\item[]\begin{tabular}[b]{@{}ll}
\br $q$&$E_0+E_1$\\
\mr
1&3.79967302980139416878309418851256895776606546733\\
2&7.455697937986738392156591347185767488137819536750\\
3&11.6447455113781620208503732813709364365508721620\\
4&16.2618260188502259378949544303846135342445865045\\
5&21.2383729182359400241497111135886363767048320597\\
20&122.604639000999455020762971417615181874976633223\\
38&284.068590581400743150496281208125064777084713267\\
39&293.948458266006085433669997483521626303445899275\\
\br
\end{tabular}
\end{indented}


\end{document}